\documentclass[twocolumn,showpacs,preprintnumbers,amsmath,amssymb]{revtex4}

\usepackage{graphicx}
\usepackage{dcolumn}
\usepackage{bm}

\def\lapproxeq{\lower .7ex\hbox{$\;\stackrel{\textstyle
<}{\sim}\;$}}
\def\gapproxeq{\lower .7ex\hbox{$\;\stackrel{\textstyle
>}{\sim}\;$}}

\begin{document}

\hyphenation{Es-ta-du-al}

\preprint{BA-02-32}

\title{Atmospheric shower fluctuations and the
 constant intensity cut method}

\author{Jaime Alvarez-Mu\~niz}
\email{alvarez@bartol.udel.edu}
\author{Ralph Engel}
\altaffiliation[Now at ]{Forschungszentrum Karlsruhe, Institut f\"ur
Kernphysik, Postfach 3640, 76021 Karlsruhe, Germany}
\author{T.K. Gaisser}
\author{Jeferson A. Ortiz}
\altaffiliation[Now at ]{Instituto de F\'{\i}sica ``Gleb Wataghin'',
Universidade Estadual de Campinas 13083-970 Campinas-SP, Brazil.
}
\author{Todor Stanev}
\affiliation{Bartol Research Institute,\\
University of Delaware, Newark, Delaware 19716, U.S.A.\\
}

\date{\bf DRAFT: August 26, 2002}


\begin{abstract}

We explore the constant intensity cut method that is widely
used for the derivation of the cosmic ray energy spectrum,
for comparisons of data obtained at different 
atmospheric depths, for measuring average shower profiles, 
and for estimates of the 
proton-air cross section from extensive air shower data. 
The constant intensity cut method is based on the selection of air
showers by charged particle or muon size and therefore is subject to
intrinsic shower fluctuations.
We demonstrate that, depending on the selection method, shower
fluctuations can strongly influence the characteristics of the selected
showers. Furthermore, a mixture of different primaries in the cosmic
ray flux complicates the interpretation of measurements based
on the method of constant intensity cuts.
As an example we consider data 
published by the Akeno Collaboration. The interpretation of the Akeno
measurements suggests that 
more than $60-70\%$ of cosmic ray primaries in the energy 
range $10^{16}-10^{17}$ eV are heavy nuclei.
Our conclusions depend only weakly on the hadronic interaction
model chosen to perform the simulations, namely SIBYLL and QGSjet. 
\end{abstract}



\maketitle

\section{\label{introduction}Introduction}

Measuring extensive air showers (EAS) is currently 
the only way to study the cosmic
ray spectrum and chemical composition at energies above 10$^{14}$ eV,
as well as the basic properties of hadronic interactions at $\sqrt{s}$
above 1.8 TeV. 

 EAS can be detected with air shower arrays which measure
 densities of shower particles such as electrons, muons, photons,
 and sometimes hadrons arriving at the detector. 
 These densities are typically fit to lateral distribution
 functions to derive the total number of charged particles,
 electrons $N_e$ and muons $N_\mu$ at detector level.
 The particle numbers are functions of the primary cosmic ray
 energy $E$ and the mass number $A$ of the primary particle, and depend
 on the atmospheric depth of the observation level.
 At energies $E$ \gapproxeq $10^{17}$ eV  the shower evolution
 can also be directly observed by measuring the fluorescence 
 light from the atmospheric nitrogen that is excited by the
 ionization of the charged shower particles.
 In the following we will concentrate on air shower arrays.
 Imaging methods such as fluorescence or Cherenkov light techniques
 will be discussed elsewhere \cite{Alvarez-Muniz02c}.

 One of the classical methods in the analysis of air
 shower data is the constant intensity cut method. The idea
 is based on the fact that, due to the isotropy of the primary
 cosmic ray flux, showers generated by primary particles
 of the same energy and composition will arrive at the 
 detector with the same frequency, assuming 100\% detection
 efficiency. Selecting showers arriving at the detector with 
 the same frequency under different zenith angles allows the
 measurement of the mean longitudinal shower profile. At large 
 atmospheric depths after shower maximum, 
 the shower size decreases approximately
 exponentially with depth with a length scale 
 commonly referred to as the {\it attenuation} length.  

 On the other hand, selecting showers with the same features
 (i.e. shower size, muon size etc.) at observation  level and
 different incident angles allows the measurement of the
 {\it absorption} length which determines how the flux of the
 selected showers decreases with atmospheric depth. 


 Measurements of the attenuation length are commonly used to correct
 observed particle densities to those of equivalent vertical showers.
 By unfolding the geometry-related attenuation of showers an
 experiment can use the measured intensities of showers with fixed size
 to derive the primary all-particle flux.
   
 The absorption length is inherently related to the mean free path of
 the EAS initiating primary particle. For example, the rate of 
 proton air showers having the first interaction point ($X_{\rm int})$ at 
 a slant 
 depth greater than $X$ decreases as $\exp(-X/\lambda_{\rm int})$,
 where $\lambda_{\rm int}$ is the mean free path for $p-{\rm air}$
 collisions. On this basis, several methods of extracting the $p-{\rm air}$ 
 cross section\footnote{The measured mean free path corresponds to
 the particle production cross section. For a discussion of the
 difference between the inelastic and production cross sections in
 relation to air showers see \protect\cite{Nikolaev93b,Engel98c}.} 
 from measurements of EAS have been applied in air shower experiments 
 \cite{Hara83,Baltrusaitis84,Honda93,Aglietta97,Aglietta99,Aglietta99b}. 

 Air shower arrays cannot measure the depth of the first interaction
 of the primary particle generating the observed shower, 
 which directly relates to the mean free path. The decrease with zenith
 angle of the frequency of showers having the same electron $N_e$ and 
 muon $N_\mu$ sizes at observation level is studied instead. 
 In the absence of intrinsic shower fluctuations
 these measurements would reflect the depth distribution of primary
 interactions. However, the longitudinal development of showers 
 is itself subject to large fluctuations. To disentangle these fluctuations
 from those of the first interaction point is not an easy task.

 This problem is usually addressed by introducing a coefficient
 ($k$) which relates the observed shower absorption length
 ($\Lambda_{\rm obs}$) and the inelastic cross section through
 the equation
 $\Lambda_{\rm int}= k\times \lambda_{\rm obs}$~\cite{Elsworthetal82}.
 The numerical value of $k$ has to be obtained from simulations
 of EAS. The coefficient $k$ reflects the influence of the
 features of the hadronic interactions model on the fluctuations
 in the shower development. The value of $k$ depends on the cross
 sections, secondary particle multiplicity and elasticity in the
 hadronic interaction model.
 Due to the necessary extrapolation of hadronic multiparticle
 production to unmeasured regions of the phase space and to high
 energy, the extracted cross section becomes model dependent.

 Further difficulties in determining the inelastic p-air cross 
 section from EAS measurements are related to experimental
 uncertainties and limitations in the determination of the
 development of  air showers, and also to the fact that the
 cosmic ray flux might be ``contaminated'' with primaries 
 heavier than protons which in principle tend to decrease the
 observed mean free path.

 In this article we shall study the importance of intrinsic shower
 fluctuations for the experimentally observed attenuation and
 absorption lengths by considering two examples:\\
 (i) the reconstruction of the primary cosmic ray spectrum using
 charged particle shower sizes, and\\
 (ii) the measurement of the proton-air cross section, following
 closely the method applied first by the Akeno group~\cite{Hara83,Honda93}
 which we call the constant $N_\mu, N_e$ method.

In the process of (ii) we found that the Akeno observations
can best be understood if there is a large fraction of heavy
nuclei present in the cosmic ray beam from 10-100 PeV.


 The paper is structured as follows. In Section~\ref{cuts} we 
 study the possible errors introduced in the derivation of the
 primary cosmic ray spectrum by shower fluctuations when
 the constant intensity method is applied. 
 Section~\ref{protons} consists of three parts. Part A 
 summarizes the basics of the constant $N_e-N_\mu$ method. In part
 B we describe the predictions of this method for proton induced
 showers and in part C we discuss the more realistic situation of
 a mixed primary cosmic ray composition.
Section~\ref{conclusions} concludes the paper.    

\section{\label{cuts} Derivation of the cosmic ray energy spectrum}

 The constant intensity cut method has been used for studies
 of the primary cosmic ray spectrum for at least 40 years.
 Let us give as an example the interpretation of the
 results of the BASJE air shower array at Mt. Chacaltaya
 performed in 1965~\cite{Bradtetal65}. 
 Since Mt. Chacaltaya is at an altitude of 5,220 m above sea
 level (540 g/cm$^2$ depth), 
 the shower size distributions obtained with the constant
 intensity cut could be used to estimate the size 
 of the showers at shower maximum. Under the assumption that
 the size at maximum is proportional to the
 primary energy $E$, this gave directly the primary energy
 spectrum within a constant which was estimated to be
 $\sim 2$ GeV/particle at shower maximum.

 In this work we apply the constant intensity cut method in
 a different way, similar to what more contemporary experiments do
 (see for instance \cite{Ave01a}).

 For illustrative purposes we first apply the method assuming
 the primary spectrum is composed of pure protons.
 We have simulated proton-induced showers at zenith angles
 $\theta=$0, 15, 30 and 45 degrees down to the altitude  
 of the Akeno array, corresponding to a vertical depth of
 $X_{\rm v}=920~{\rm g/cm^2}$.
 Shower energies were drawn from an $E^{-3}$ differential
 injection spectrum in the energy range between $10^{16}$ and
 $10^{18}$ eV. 

 We used a hybrid air shower simulation program to generate
 large samples of showers in an efficient and fast manner~\cite{Alvarez02}.
 The hybrid method consists of calculating shower
 observables by a direct simulation of the initial part of
 the shower, tracking all particles of energy above $E_{\rm thr}=0.01~E$.
 Presimulated showers for all subthreshold particles are
 then superimposed after their first interaction point is
 simulated. The subshowers are described with parametrizations
 that give the correct average behavior, and at the same time
 describe the fluctuations in shower development of both electrons
 and muons.

\begin{table}
\caption{Results of the fit to the relation between shower
energy and size at observation level for vertical showers 
at different depths $X_{\rm v}$
and assuming different cosmic ray primaries. The last row 
corresponds to the energy-shower size relation obtained for
a mixed composition assuming equal fractions of p, He, CNO and Fe.
The function used to do the fit is 
$E={\hat E} {(N_e^0)}^{(1-\epsilon)}$. \label{table:E-Ne}}
\renewcommand{\arraystretch}{1.5}
\begin{tabular}{c|c|c|c} \hline \hline
 
~Primary~&~$X_{\rm v}~[{\rm g/cm^2}$]~&~$\log_{10}({\hat E}/{\rm eV})$~&~$\epsilon$~\\ \hline
 
~p~~&~700~&~9.29~&~~0.019~~\\
 
~p~~&~870~&~9.81~&~~0.072~~\\ 
 
~p~~&~920~&~9.91~&~~0.077~~\\
 
~He~~&~920~&~10.07~&~~0.089~~\\ 

~CNO~~&~920~&~10.24~&~~0.103~~\\ 

~Fe~~&~920~&~10.63~&~~0.143~~\\ 

~mixed~~&~920~&~10.21~&~~0.103~~\\ \hline \hline

\end{tabular}
\end{table}

 The procedure we use to reconstruct the primary spectrum is the
 following.  First, from the simulations we obtain the relation
 between shower energy ($E$) and shower size at observation level 
 in vertical showers ($N_e^0$). This relation is given in
 Table~\ref{table:E-Ne}. Our simulations predict the shower size
 spectra at Akeno  level for different zenith angles, and we treat
 them as if they were actual experimental data, replacing the
 detector induced fluctuations in $\log_{10}N_e$ by a Gaussian
 resolution function of width $\Delta\log_{10}N_e=0.05$. Since we
 do not simulate showers  of energy below 10$^{16}$ eV there is an
 artificial
 break in the size spectra at low energy. To avoid it we choose
 $N_e=10^7$ as a threshold value above which our ``array'' is
 fully efficient, and we only deal with showers having $N_e$
 above this value.

 We apply cuts at constant shower intensity, and by studying
 the decrease with zenith angle of the size corresponding to each 
 intensity, we obtain the shower attenuation length, which
 we use to estimate the shower size at $\theta=0^\circ$ 
 from the known size at zenith angle $\theta$. 
 (We checked that the attenuation length obtained from 
 the simulated data in this way agrees with the attenuation
 length of the averaged profile of the input showers.)
 This is the
 classical integral application of the method. Given 
 $N_e^0$ we can use the previously obtained relation between
 $E$ and $N_e^0$ to estimate the energy of each individual
 shower. The energy spectrum for different zenith angles can
 then be reconstructed and compared to the injected spectrum. 

\begin{figure}
\centerline{
\includegraphics[width=8.5cm]{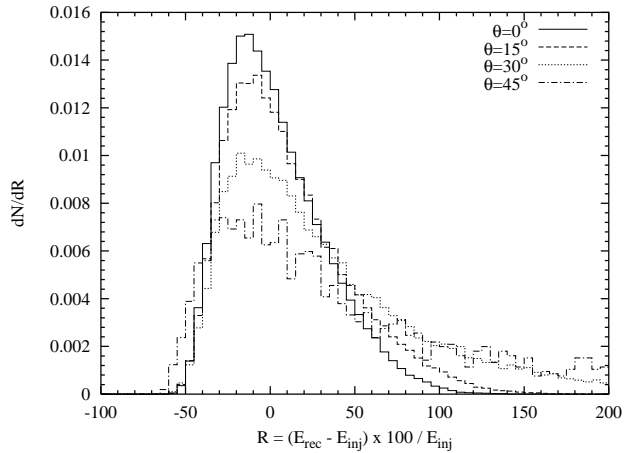}
}
\caption{Energy resolution of proton-induced showers
for different zenith angles.
The SIBYLL 2.1 hadronic generator code was used to simulate
the showers.
}
\label{eresolution}
\end{figure}

 Fig.~\ref{eresolution} shows the shower energy resolution we
 achieve with this procedure. The distribution of differences
 between reconstructed and injected energies is highly asymmetric.
 The asymmetry depends on zenith angle as showers at larger
 zenith angles are further away from their maximum where the 
 fluctuations in shower size are smallest. There is a clear
  tendency to misreconstruct showers of a certain injected energy
 assigning them a higher energy. 

\begin{figure}
\centerline{
\includegraphics[width=8.5cm]{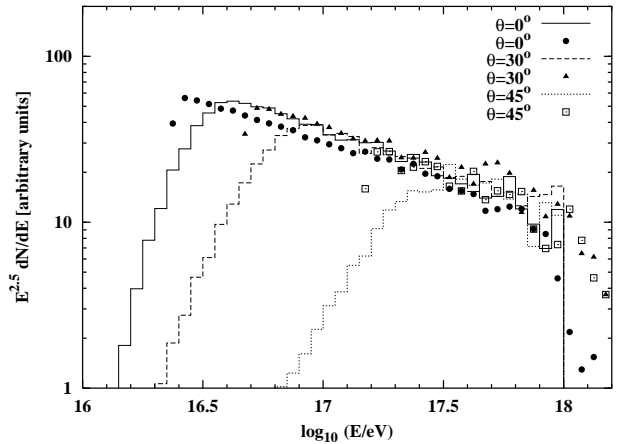}
}
\caption{Reconstructed (points) and injected (histograms) 
 energy spectra (multiplied by \protect$E^{2.5}$ at different
 zenith angles. Note that larger zenith angles contribute to
 the derived cosmic ray spectrum
 in a limited energy range because of the $N_e>10^7$ threshold.
}
\label{recons-spectr}
\end{figure}

 In Fig.~\ref{recons-spectr} we compare the reconstructed and
 injected spectra for different zenith angles.
 The spectra are multiplied by $E^{2.5}$ for a better resolution.
 Although we draw showers from an $E^{-3}$ differential spectrum,
 the cut in $N_e$ decreases the contribution of the lower energy
 cosmic rays and creates the turnaround seen in Fig.~\ref{recons-spectr}.
 The larger the zenith angle, the higher the shower energy must be 
 to exceed the $N_e$ cut, producing a strongly zenith angle dependent
 energy cut.

 Besides, as mentioned previously, with increasing zenith angle the
 shower is sampled further away from shower maximum and the fluctuations
 in $N_e$ grow, producing a broadening of the injected spectrum.
 These two effects, however, vanish almost completely in the reconstructed
 spectra.  

 Imposing a cut in $N_e$ produces a corresponding cut in $E$ through
 the relation between energy and size at observation level. 
 In our case $N_e>10^7$ implies that $\log_{10}E>16.36,~16.66,~17.16$ at
 zenith angle 0, 30 and 45 deg respectively. As a consequence all the low 
 energy injected events are reconstructed with energies above these values,
 and they pile-up rendering a simple power law. A fit to the reconstructed
 spectra reveals that the differential spectral index decreases
 by only about $2-3\%$. This is an important effect which should be
 present in experiments
 reconstructing the spectrum which use a relation between 
 shower energy and shower size at observation depth, for instance
 at a certain distance from the core of the shower. 

 Although the spectral shape is preserved almost completely
 there is a slight, but noticeable difference in the absolute
 normalization of the spectra reconstructed from showers 
 at different zenith angles. The energies derived from showers
 at non vertical angle are always overestimated, which leeds
 in principle to an artificially increased normalization.
 Fig.~\ref{recons-spectr} shows that increase for an angle of 30$^\circ$,
 however the normalization from the 45$^\circ$ showers is again lower.
 This may be due to the energy resolution distribution that peaks
 well below 0 for these showers. Independently of the exact reason
 for the changing normalization, Fig.~\ref{recons-spectr} demonstrates
 that the exact energy derivation depends strongly on the shape of
 the shower fluctuations. Since these fluctuations change with
 the atmospheric depth, so does the reconstruction accuracy.
 As a whole, though, the method works quite well. 

 The reconstruction of the shower energy from $N_e$ is affected
 even less by the shower fluctuations if the depth of the detector
 is close to the depth of shower maximum. In this case the danger
 is in the inclusion of showers that have not yet achieved their maximum
 development. Such showers may introduce a significant bias when 
 their size is converted to vertical size by using the attenuation
 length. In the example discussed above, the mean depth of shower
 maximum is $\sim 650~{\rm g/cm^2}$ with a standard deviation of
 $\sim 70~{\rm g/cm^2}$, so that only a small fraction $\sim 0.5\%$
 of the vertical showers have their maxima below observation level.
 This fraction is much smaller for inclined showers.    

 Further difficulties in the reconstruction arise in the
 more realistic case when the primary spectrum consists of a
 mixed composition of different nuclei. The heavier the primary
 nucleus, the further away is the observation level from
 shower maximum. From this point of view the spectrum reconstruction
 for a mixed primary composition is analogous to using proton showers
 to very large zenith angles.
 
 To  explore this point we have simulated a primary composition
 consisting of equal fractions of protons, He, CNO and Fe. We
 obtained the $E-N_e$ relation at $920~{\rm g/cm^2}$ (shown in
 table~\ref{table:E-Ne}) from this particular mixture of nuclei,
 and applied the same procedure as before to reconstruct the 
 primary spectrum. The reconstruction is again affected by the
 cut in $N_e$ as explained above.  
 The result is that both the spectral index and the normalization 
of the reconstructed
 spectrum differ from the corresponding values in the injection
 spectrum by only a few percent. 

 It is important to note that we have made use of our prior
 knowledge of the injected primary composition to reconstruct
 the spectrum. We expect the reconstruction method to work either 
 when the primary composition is known or when a composition
 independent energy estimator is used. (The density at 600 m from 
 the cores of large showers is an example of a measure of shower energy
 chosen because of its relative insensitivity to primary 
 mass~\cite{NaganoWatson00}.)
 We have checked that using
 the $E-N_e$ relation obtained for pure protons tends to underestimate
 the normalization of a mixed spectrum. The reason is that a shower
 of energy $E$ initiated by a heavy nucleus has on average a
 smaller size at observation level than a proton shower of the 
 same energy.  A smaller energy is then assigned to the shower when 
 the $E-N_e$ relation for protons is used. As expected, we observed
 the opposite behavior when the $E-N_e$ relation for pure iron
 (also shown in table~\ref{table:E-Ne}) is used to reconstruct the
 mixed composition spectrum.    

 The most difficult case is obviously that of changing chemical
 composition. Because of the changes of the spectrum normalization
 for different primary nuclei the shape of the spectrum can also
 be derived incorrectly. The composition and spectrum then have
 to be reconstructed simultaneously from different shower parameters. 

 We conclude that the integral application of the constant
 intensity cuts method for the derivation of the primary
 cosmic ray spectrum works well when the cosmic ray composition
 is known.  
The use of wrong composition models can lead to
 erroneous conclusions for the energy spectrum, mostly in
 the determination of its normalization. 
The method is not strongly affected by the intrinsic
 fluctuations in the shower in the energy range we have explored.

\section{\label{protons} The constant $N_e-N_\mu$ method}  


 The total column density of atmosphere available for
 shower development increases with the incident angle $\theta$ 
 as $\sec\theta$.
 The total number of electrons at a fixed slant depth after
 the shower maximum reflects the stage of evolution of the
 shower. If shower fluctuations were absent, selecting showers
 of fixed energy at different zenith angles which have the same
 electron size $N_e$, would a priori guarantee that they have developed
 through the same column of atmosphere between the first interaction
 point and observation level.
 The selected showers would only differ in the depth at which the
 first primary p-air interaction had occurred. The proton-air
 interaction length $\lambda_{p-air}$ and the corresponding
 cross section $\sigma_{p-air}$ would then be measured.
 The fact is, however, that $N_e$ does have large fluctuations.
 
 To address this problem and select showers of fixed primary energy,  
 experiments
 often require that the showers have the same muon size at
 observation level $N_\mu$. Unlike electrons, the number of muons $N_\mu$
 remains almost constant after maximum, and hence
 it is a good estimator of the primary energy at essentially any
 observation depth below shower maximum.
 Selecting the showers with large electron sizes within the
 same $N_\mu$ bin increases the probability that they are induced
 by protons. The higher the size is, the lower is the contamination
 from heavier primaries.  
 
 Once showers are selected in this way, the frequency ($f$) 
 of showers falling in a given ($N_\mu,N_e$) bin is measured for 
 different zenith angles. 
 The ratio of the frequency of selected showers
 at two zenith angles ($\theta_1$ and $\theta_2$) 
 is related to the observed absorption length by
\begin{equation}
R(\theta_1,\theta_2)=\frac{f(N_\mu,N_e,\theta_1)}
{f(N_\mu,N_e,\theta_2)} =
\exp{\left[-\frac{X_{\rm v}}{\Lambda_{\rm obs}}
(\sec\theta_1 - \sec\theta_2)\right ]}\ ,
\label{eq:ratio}
\end{equation}
where $X_{\rm v}$ is the vertical depth of the detector.

 Clearly, intrinsic fluctuations of the shower profile will
 change the relation between the first interaction point and 
 the electron and muon number at larger depths. In the following we will
 study how such fluctuations influence the observed absorption 
 length, using detailed, up-to-date hadronic interactions models.
 For definiteness we will concentrate on the implementation of the
 method as used by the Akeno group~\cite{Hara83,Honda93}. A similar
 procedure was also used more recently
 by the EAS-TOP Collaboration \cite{Aglietta99b}.

\subsection{Application to proton-induced showers}

 In general, the primary cosmic ray flux consists of 
 nuclei of a variety of mass numbers. Since we 
 want to study the constant $N_e-N_\mu$ method itself we simplify 
 the problem and start with the assumption 
 that all primary particles are protons. If the procedure does not give
 the correct cross section for a purely proton
 flux, the correct derivation for a mixed cosmic ray
 composition would be impossible.

 We have performed simulations of proton-induced showers 
 at several zenith angles and calculated the frequency of showers
 having $N_\mu$ muons and $N_e$ electrons at observation level. 
 The detector was chosen to be located at Akeno altitude, 
 corresponding to a vertical depth of $X_{\rm v}=920~{\rm g/cm^2}$.
 Shower energies were drawn from an $E^{-3}$ differential 
 spectrum in the energy range between $10^{16}$ and 
 $10^{18}$ eV. We performed the simulations for fixed
 zenith angles ($\theta=$0, 15, 30 and 45 deg.). We thus
 simplify the problem once again by neglecting 
 the errors introduced by the experimental 
 shower zenith angle reconstruction.
 
 To study the dependence on the hadronic interaction model
 we have performed our simulations with two models, namely SIBYLL 2.1 
 \cite{Engel99,Engel01} and QGSjet98 \cite{qgsjet}. The two
 models give similar predictions for the shower development in   
 the energy range $10^{16}-10^{18}$ eV \cite{Alvarez02}. As will be 
 shown, our results depend only weakly on the choice of model.  

 For a realistic simulation of the observed shower parameters
 one should account for the experimental uncertainty and
 fluctuations due to the detector. A detailed simulation of the
 biases and efficiencies of the detectors is beyond the
 scope of this paper. It requires the use of specifically designed
 Monte Carlo programs of each particular ground array. We replace
 instead the detector induced fluctuations in $\log_{10}N_\mu$ and
 $\log_{10}N_e$ by Gaussian resolution functions of widths
 $\Delta \log_{10}N_\mu=0.1$ and $\Delta \log_{10}N_e=0.05$
 respectively. These are the experimental errors reported by the
 Akeno group \cite{Honda93}. For each of the simulated showers
 modified $\log_{10}N_\mu$ and $\log_{10}N_e$ are sampled according
 to the theoretical values and the detector resolution.

It is possible and even likely that, due to different energy
 thresholds and absorbing materials, the experimental
 definition of $N_e$ does not coincide exactly with the corresponding
 quantity that the Monte Carlo generates. In such a case one should
 apply a correction factor to achieve a full reproduction of the
 experimental result. As we show further below, however,
 a possible small discrepancy in the definition of $N_e$ would
 not alter the conclusions of this study.

We have simulated a sample of 500,000 proton showers, 
comparable to the statistics of the full event sample
reported by the Akeno collaboration 
in \cite{Honda93}.

\begin{figure}
\centerline{
\includegraphics[width=8.5cm]{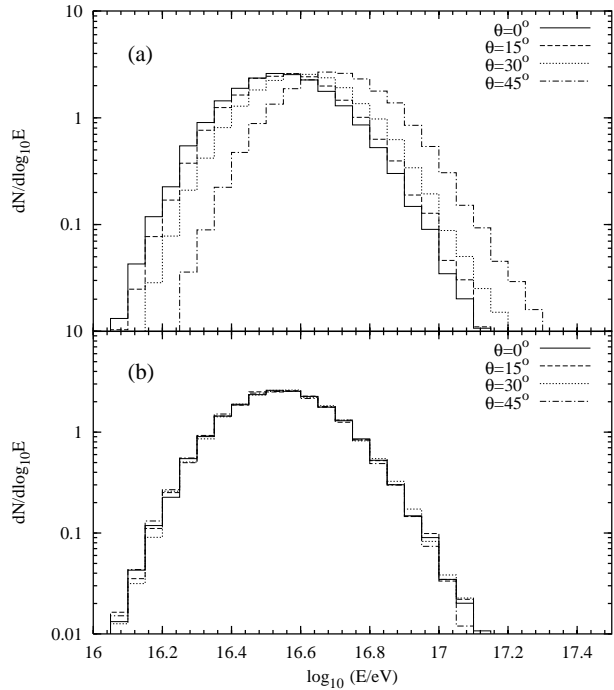}
}
\caption{Top panel: Simulated energy distribution of
proton-initiated showers after
applying the muon cut. The selected showers have
$\log_{10} N_\mu$ between 5.25 and 5.45.
The muons have energy above $E^{\rm thr}_{\mu}=1~{\rm GeV} \times \sec\theta$
at 920 ${\rm g/cm^2}$. The distribution is shown for different zenith
angles. The SIBYLL 2.1 hadronic generator code was used to 
simulate the showers. Bottom panel: Same as top panel after applying the
constant intensity cut.
}
\label{edistr}
\end{figure}
 We apply the constant $N_e-N_\mu$ method by first selecting
 showers which have $\log_{10}N_\mu$ between 
 5.25 and 5.45 at observation level, the muon number
 of the first bin of the Akeno analysis \cite{Honda93}. 
Only muons with energy $E_\mu>1~{\rm GeV}\times \sec\theta$,
which is the muon energy threshold of the Akeno
experiment~\cite{Honda93},
 are considered.  
 In the top panel of Fig.~\ref{edistr} we plot the energy
 distribution of showers selected after applying the muon cut. 
 For fixed primary energy the mean muon number is smaller at large
 zenith angles as compared to vertical showers.
 This is due both to the dependence of the energy threshold
 on zenith angle, and to the increase of the probability for muon
 decay when the zenith angle increases and muons have to
 traverse more column depth of atmosphere.
 As a consequence, the energy distribution shows a dependence
 on zenith angle, i.e. selecting showers with the same number
 of muons does not perfectly guarantee that they have the 
 same energy distribution. One can correct for the $N_\mu$
 attenuation by the constant intensity cut method 
 shifting slightly the $\log_{10}N_\mu$ bin so that the intensity
 of showers is the same at all zenith angles.
 This is equivalent to a correction of the shower muon longitudinal
 profile as a function of the zenith angle. 
 The bottom panel of Fig.~\ref{edistr} shows the energy 
 distribution of the selected showers after applying the 
 constant intensity cut method. The constant intensity cut 
 method works almost perfectly, giving a distribution of
 selected shower  energies independent of zenith angle.    
 The width of the shower energy distribution is determined by
 the width of the $N_\mu$ bin, and by the $N_\mu$ 
 shower to shower fluctuations.

\begin{figure}
\centerline{
\includegraphics[width=8.5cm]{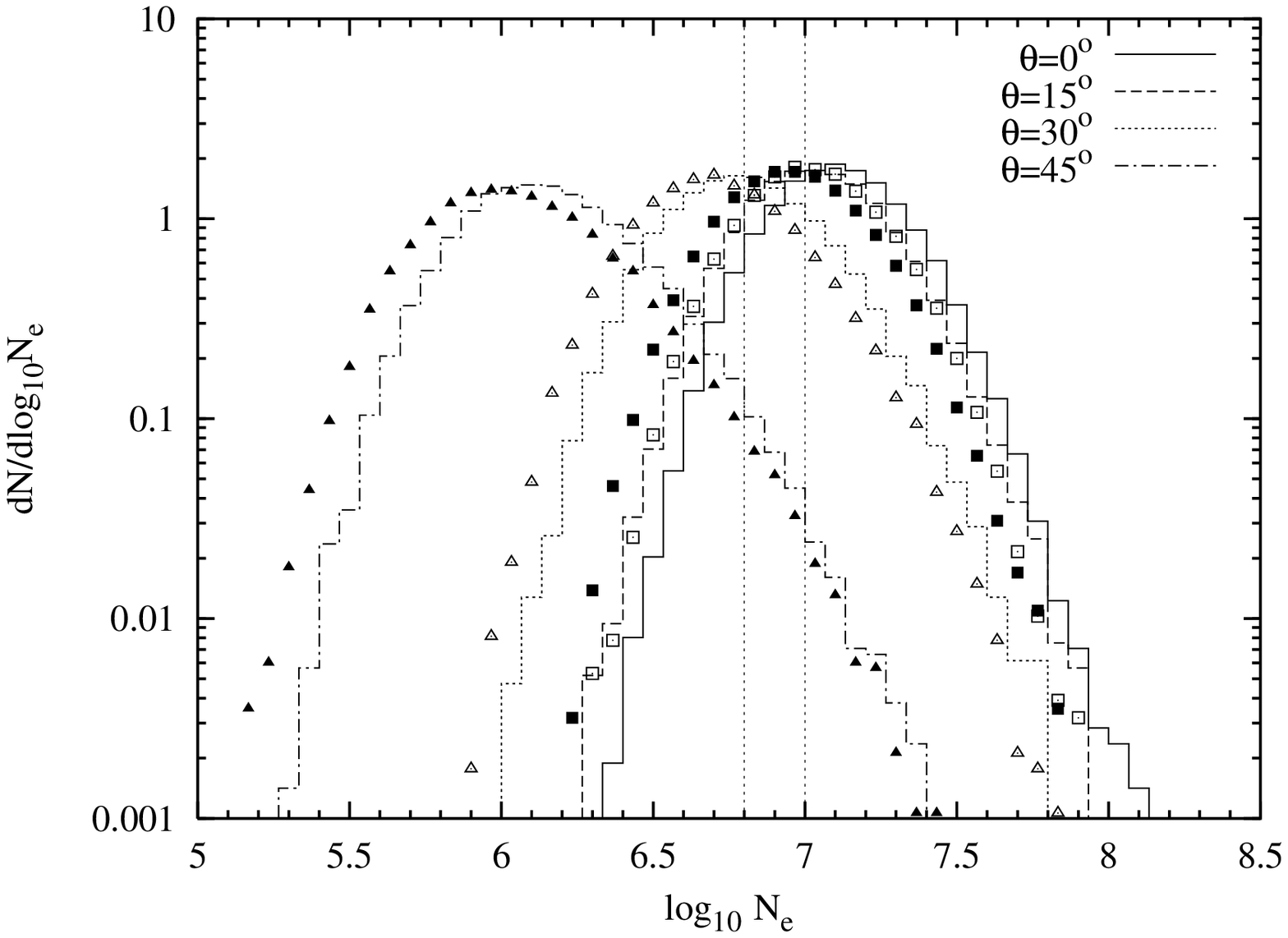}
}
\centerline{
\includegraphics[width=8.5cm]{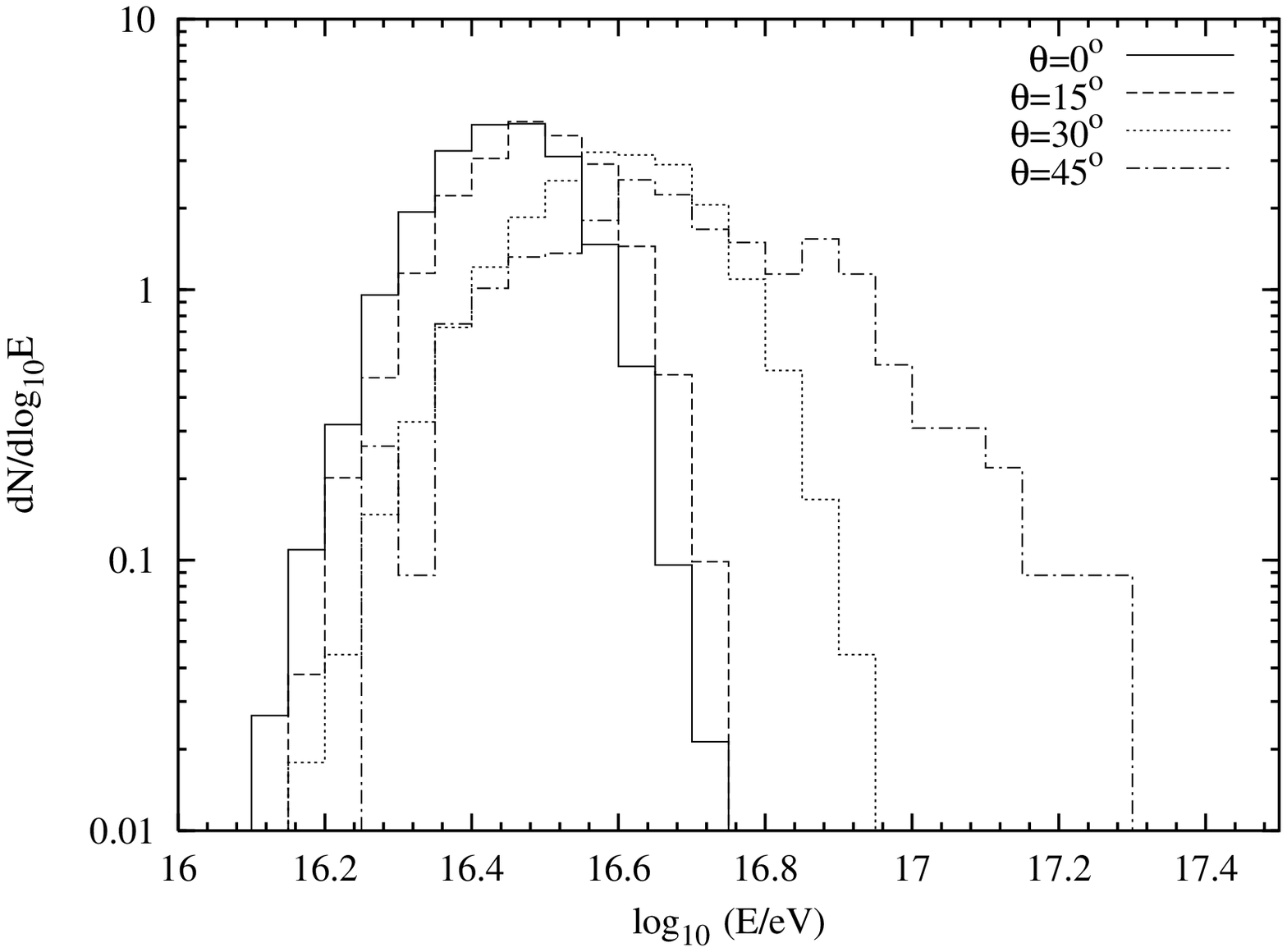}
} 
\caption{
Top panel: Shower size at 920 ${\rm g/cm^2}$ depth in 
proton-initiated showers having between $10^{5.25}$ and 
$10^{5.45}$ muons at 920 ${\rm g/cm^2}$. 
The size distribution is shown for showers initiated at
different zenith angles. 
Histograms correspond to showers simulated using SIBYLL 2.1,
and points to showers simulated with QGSjet98. Bottom panel:
Energy distribution of the showers falling in the \protect$N_e$
 bin indicated by the vertical bars in the top panel.
}
\label{sobs}
\end{figure}

 Once showers of the same energy have been selected, we select
 in addition showers with constant $N_e$, as was done by the
 Akeno team.
 The top panel of Fig.~\ref{sobs} shows the $N_e$ spectra of
 showers having $\log_{10}N_\mu$ between 5.25 and 5.45 for the
 four nominal zenith angles. 
 The two vertical lines mark the bin in $\log_{10}N_e$ 
 chosen by the Akeno collaboration to perform their analysis
 of the showers in the 5.25-5.45 bin in $\log_{10}N_\mu$.
 The bottom panel of Fig.~\ref{sobs} shows the energy distribution
 of showers in the selected $N_e$ bin. It shows that the 
 energy estimate that was very good after the $N_\mu$ bin
 selection with constant intensity cuts, is now again
 angle dependent.

 We have compared the $N_e (E)$ dependence calculated by
 the Monte Carlo code to the experimental one used by the 
 Akeno experiment \cite{Nagano84}.
 In the $\log_{10} N_e$ bin 6.8 - 7.0 the Akeno formula
 gives $\log_{10}(E/{\rm eV})$ = 16.40 and the Monte Carlo code
 using SIBYLL 2.1 yields 16.43. Both values are estimated at 
 the center of the bin. The differences when using the $N_\mu (E)$ 
 are higher.
 The Akeno collaboration~\cite{Nagano84} derived $\log_{10}(E/{\rm eV})$
 = 16.25 in the $\log_{10} N_\mu$ bin 5.25 - 5.45.
 SIBYLL 2.1 gives $\log_{10}(E/{\rm eV})=16.55$ and QGSjet98 
 $\log_{10}(E/{\rm eV})=16.48$ in the same $\log_{10} N_\mu$ bin. 

\begin{figure}
\centerline{
\includegraphics[width=8.5cm]{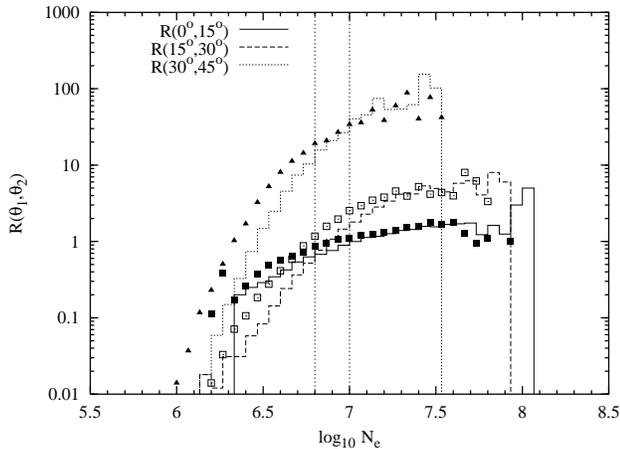}
}
\caption{
Ratios of number of proton-initiated showers having between 
$10^{5.25}$ and $10^{5.45}$ muons and electron size
$N_e$ at 920 ${\rm g/cm^2}$ as a function of $N_e$. 
Histograms correspond to showers simulated using SIBYLL 2.1,
and points to showers simulated with QGSjet98.
}
\label{ratio}
\end{figure}

 In Fig.~\ref{ratio} we further illustrate this dependence by 
 showing the frequency ratios of the showers
 in Fig.~\ref{sobs}, in dependence of the selected electron size. 
The ratio is only plotted for
adjacent zenith angles. It depends strongly on the $N_e$
bin used for shower selection. According to Eq.~(\ref{eq:ratio})
this ratio should be constant over a certain range in $N_e$
for all different zenith angle combinations.
Fig.~\ref{ratio} shows that the $N_e$ ranges where the ratio is
approximately constant depend on the shower angle combination. 
For both SIBYLL 2.1 and
QGSjet98 models they lie above $\log_{10} N_e$ of 7.4
for the $\log_{10} N_\mu$ bin 5.25-5.45. 
The bin in $N_e$ chosen by Akeno for the cross section analysis is
clearly in a region where the intensity ratios depend strongly on $N_e$.

The ratio of two ratios $R$ for different combinations
of zenith angles can be used as a consistency check of
the results. The expected values of the double ratio are (see
Eq.~(\ref{eq:ratio}))
\begin{eqnarray}
\frac{\log_{10}R(15^\circ,30^\circ)}{\log_{10}R(0^\circ,15^\circ)}=
\frac{\sec(30)-\sec(15)}{\sec(15)-\sec(0)}\sim 3.4\\[0.5cm] 
\frac{\log_{10}R(30^\circ,45^\circ)}{\log_{10}R(0^\circ,15^\circ)}=
\frac{\sec(45)-\sec(30)}{\sec(15)-\sec(0)}\sim 7.4
\end{eqnarray}
It is easy to verify that this is not the case for the selected
showers falling in the bin in $\log_{10}N_e$ between 6.8 and 7.0.
These numbers are, however, the approximate scaling
factors when comparing the plateaus of the ratios in
Fig.~\ref{ratio}. The figure suggests that zenith angle
dependent bins in electron size should be used in order
to get an angular-independent value of $\Lambda_{\rm obs}$,
provided of course the selected showers are initiated by protons
as in our simulation.

\begin{figure}
\centerline{
\includegraphics[width=8.5cm]
{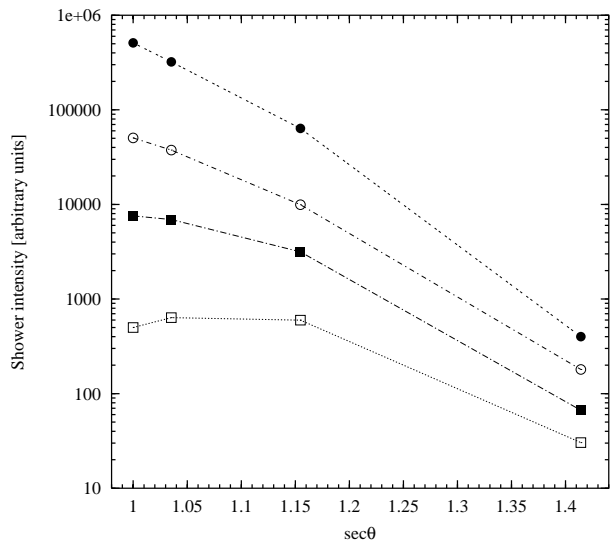}
}
\caption{Zenith angle dependence of the intensity of proton-induced showers
having constant $\log_{10}N_\mu=5.25-5.45$ and constant
$\log_{10}N_e$ for different values of $\log_{10}N_e$. 
Empty squares $\log_{10}N_e=6.8-7.0$, 
filled squares $\log_{10}N_e=7.0-7.2$, empty circles 
$\log_{10}N_e=7.2-7.4$ and filled circles $\log_{10}N_e=7.4-7.6$. 
Showers were simulated
with SIBYLL 2.1. The points are joined by straight
lines to guide the eye. To avoid overlapping, the results
for different $N_e$ bins were multiplied by different
arbitrary factors.
}
\label{zenith_ne_cuts}
\end{figure}

 A similar consistency check can be performed by plotting the
 observed shower intensity as a function  of $\sec \theta$.
 Fig.~\ref{zenith_ne_cuts} shows the intensity of proton-initiated
 showers falling in the $\log_{10}N_\mu=5.25-5.45$
 bin for different $N_e$ bins. The deviation from straight
 lines, which are expected for exponential attenuation of
 showers with $\sec \theta$, demonstrates that the constant
 $N_e-N_\mu$ method fails to select similar showers, unless
 a large value of $N_e$ is selected. 
 The intensity of the selected showers certainly does not decrease
 as $\exp(-X/\Lambda_{\rm obs})$ in the  
 $(\log_{10}N_\mu,\log_{10}N_e)=(5.25-5.45,6.8-7.0)$ bin.
Formal fits of the three higher $N_e$ bins plotted in
Fig.~\ref{zenith_ne_cuts} give $\Lambda_{\rm obs}$ values of
$138 \pm 36$, $85 \pm 8$ and $68 \pm 3$ g/cm$^2$ for $\log_{10} N_e$
of 7.0-7.2, 7.2-7.4 and 7.4-7.6 respectively. Compared to
the proton-air cross section of 456 mb in SIBYLL 2.1
these values lead to $k$ values of $2.60 \pm 0.67$, $1.61 \pm 0.14$,
and $1.28 \pm 0.06$.  

The analogous analysis carried out with QGSjet98 for the 
same muon number bin and $\log_{10} N_e = 7.4-7.6$ 
gives $\Lambda_{\rm obs} = 69 \pm 2$ which corresponds 
to a $k$-factor of $1.26 \pm 0.04$. Within the statistical uncertainty
this value agrees with the one derived from SIBYLL 2.1 simulations.
The weak model dependence is not unexpected. The energy of the showers
considered here is only one order of magnitude higher than the
equivalent energy of the Tevatron collider. Both models were tuned to
reproduce the Tevatron measurements and predict rather similar muon and
electron numbers for $E \ll 10^{19}$ eV. 

\subsection{Shower fluctuations}

\begin{figure}
\centerline{
\includegraphics[width=8.5cm]{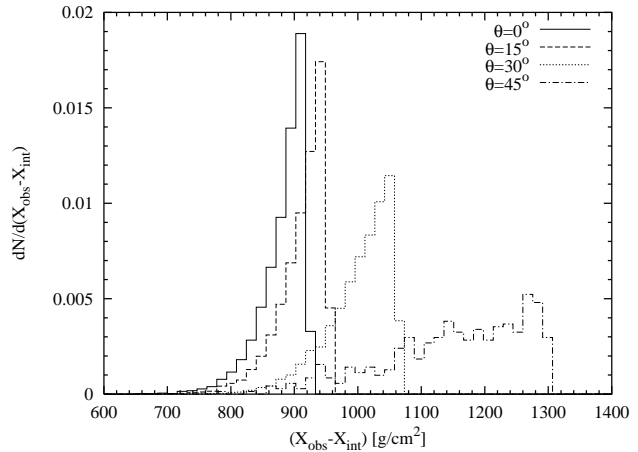}
}
\caption{Distribution in $X_{\rm obs}-X_{\rm int}$  
of the showers that fall 
in the ($\log_{10}N_\mu,\log_{10}N_e$)=(5.25-5.45,6.8-7.0) bin. 
}
\label{xobs-xint}
\end{figure}

The ultimate reason why the constant
$N_e-N_\mu$ method does not work 
is that the discussed shower selection is dominated by the 
intrinsic fluctuations in shower development. 
 This is illustrated in Fig.~\ref{xobs-xint} in which we 
 plot the distribution of ``shower lengths'' of showers with
 $(\log_{10}N_\mu,\log_{10}N_e)=(5.25-5.45,6.8-7.0)$ for different angles. 
 We arbitrarily define the shower length as the difference between
 the slant depth of observation level ($X_{\rm obs}$) 
 and the slant depth of the first interaction point. If the
 longitudinal shower profile were not biased by the selection
 criteria all four histograms would be very similar.
 They are instead very different and demostrate
 that the selection is on the width of shower development
 rather than on the depth of the first interaction $X_{\rm int}$.
 Indeed, for all angles most of the selected showers have
 their first interaction point near the top of the atmosphere.
 Even in the larger $N_e$ bins the situation is not qualitatively
 different, as can be seen in Fig.~\ref{zenith_ne_cuts}.

In table~\ref{shfluct} we give the average values and the
widths of the distributions shown in Fig~\ref{xobs-xint}.
For the method not to be dominated by intrinsic fluctuations,
the average value of $X_{\rm obs}-X_{\rm int}$ should be 
independent of zenith-angle. The increase of atmospheric
depth from $\theta=0^\circ$ to 45$^\circ$ should lead to a shift
of the first interaction point by about 400 g/cm$^2$. Due to the
fluctuations the actual mean shift is only by about 110 g/cm$^2$.

 A way to quantify which part of the longitudinal shower
 development, namely the first few interactions or the
 latest interactions, contributes most to the fluctuation
 in shower length, is calculating the average values and widths of the
 distributions of $X_{\rm max}-X_{\rm int}$ and 
 $X_{\rm obs}-X_{\rm max}$. 
 This is shown in columns 3 and 4 in the table. The tail of the
 shower contributes more to the overall fluctuation in shower
 length than the first few interactions, although the contribution
 depends on the zenith angle. In terms of the ratio of $\sigma/\Delta X$
 the angular dependence is bigger for $(X_{\rm obs} - X_{\rm max})$,
 where it changes from 0.054 for vertical showers to 0.136 for
 showers developing under 45$^\circ$.
\begin{table*}
\caption{Average values and standard deviations (in parenthesis) 
of the distributions of $X_{\rm obs}-X_{\rm int}$ 
(see also Fig.~\ref{xobs-xint}), 
$X_{\rm max}-X_{\rm int}$ and $X_{\rm obs}-X_{\rm max}$
for proton-initiated showers belonging to the 
($\log_{10}N_\mu,\log_{10}N_e$)=(5.25-5.45,6.8-7.0) bin.
\label{shfluct}}
\renewcommand{\arraystretch}{1.5}
\begin{tabular}{c|c|c|c|c} \hline \hline
 
~$\theta$~[deg.]~&~$X_{\rm obs}~[{\rm g/cm^2}]$&~$(X_{\rm obs}-X_{\rm int})~[{\rm g/cm^2}]$~&~$(X_{\rm max}-X_{\rm int})~[{\rm g/cm^2}]$~&~$(X_{\rm obs}-X_{\rm max})~[{\rm g/cm^2}]$~\\ \hline
 
~0~&~920.0&~881.3~~(35.7)~&~581.4~~(31.1)~&~300.1~~(42.6)~\\ \hline

~15~&~952.2&~911.1~~(37.6)~&~585.8~~(31.6)~&~325.2~~(42.3)~\\ \hline

~30~&~1062.3&~1002.0~~(50.6)~&~613.7~~(39.5)~&~388.2~~(51.3)~\\ \hline

~45~&~1301.1&~1152.9~~(109.7)~&~699.7~~(95.1)~&~453.8~~(92.5)~\\ \hline\hline
 
\end{tabular}
\end{table*}

\subsection{\label{composition}Composition}

 In contrast to our findings summarized in Fig.~\ref{zenith_ne_cuts},
 the Akeno Collaboration reports a $\sec \theta$ dependence of the
 observed frequencies of showers selected by the constant $N_e-N_\mu$
 method which is compatible with an exponential attenuation (see
 Fig.~1 in Ref.~\cite{Honda93}).
 By looking at Figs.~\ref{sobs} and \ref{zenith_ne_cuts} it is clear that
 the intensity of proton showers in the nominal bin is not large enough at
 small zenith angles to produce a straight line.  Proton showers
 penetrate too much in the atmosphere and 
 thus have large electron sizes at observation level. 
 In principle, this statement depends on the hadronic interaction model
 used in the simulations. A model which predicts the same muon number at
 lower shower energy can lead to an increase of the vertical shower 
 intensity in the considered bin.
However, simulations with QGSjet, which predicts the largest muon 
multiplicity among the contemporary hadronic interaction models, 
show that this conclusion is unchanged if the muon multiplicity in the
considered energy range is increased by up to 20\%.

\begin{figure}
\centerline{
\includegraphics[width=8.5cm]{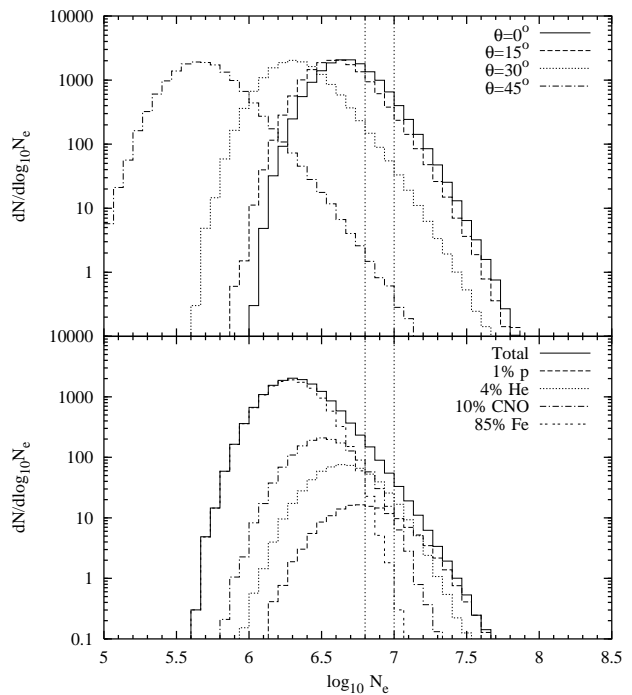}
}
\caption{
Top panel: Same as Fig.~\ref{sobs} for a primary cosmic ray composition 
consisting of $85\%$ Fe, $10\%$ CNO, $4\%$ He and $1\%$ protons.
The bottom panel shows the $\theta=30^o$ size distribution that
is plotted in the top panel illustrating how the different cosmic 
ray primaries contribute to build it up.
}
\label{sobs_compos}
\end{figure}

 A way to increase the intensity of the selected showers would
 be to  ``contaminate'' the sample with heavy primaries.
 These give rise to showers which are less penetrating, shifting
 the distribution in electron size to smaller values. This is
 illustrated in the top panel of Fig.~\ref{sobs_compos} for a
 primary composition consisting of $85\%$ Fe, $10\%$ CNO, $4\%$ He 
 and $1\%$ protons. The bottom panel of Fig.~\ref{sobs_compos} 
 shows the contributions to the total electron size distribution
 from showers initiated by the different primaries. It is 
 remarkable, and to our understanding a coincidence, that the tail of 
 the total distribution in electron size has roughly the same slope
 as the tail of the contribution from proton-induced showers alone.  

\begin{figure}
\centerline{
\includegraphics[width=8.5cm]
{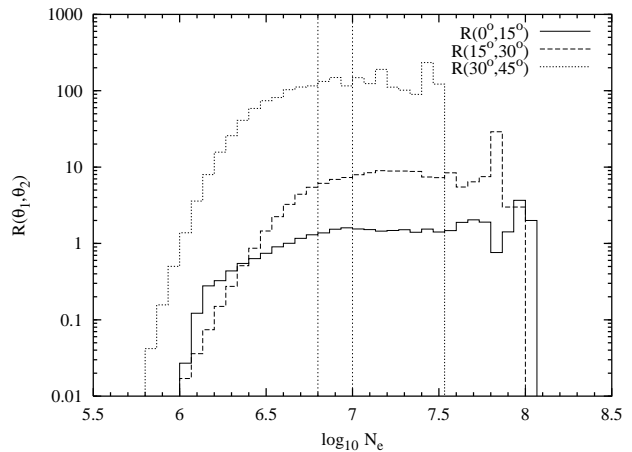}
}
\caption{
Ratios of number of proton-initiated showers having between
$10^{5.25}$ and $10^{5.45}$ muons and electron size
$N_e$ at 920 ${\rm g/cm^2}$ as a function of $N_e$
for a primary cosmic ray spectrum consisting on
$85\%$ Fe, $10\%$ CNO, $4\%$ He and $1\%$ protons.
Showers were simulated with SIBYLL 2.1.
}
\label{ratio_compos}
\end{figure}

 Fig.~\ref{ratio_compos} shows the ratio of the electron 
 size distributions shown in the top panel of Fig.~\ref{sobs_compos} 
 for adjacent zenith angles. In the $N_e$ range where 
 the ratios are flat, they have numerical values very similar  
 to the expected ratios for protons shown in Fig.~\ref{ratio}
 in the corresponding plateau regions, i.e. although the primary 
 spectrum is dominated by heavy primaries the analysis method gives
 a value of the cross section similar to, but somewhat 
 higher than that obtained for pure protons. 

\begin{figure}
\centerline{
\includegraphics[width=8.5cm]
{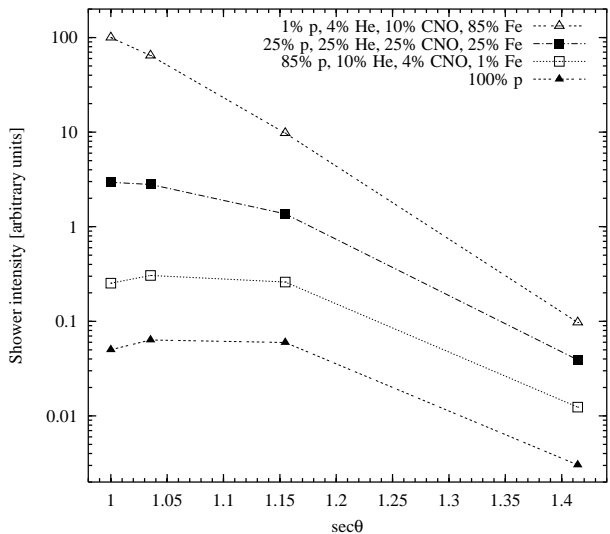}
}
\caption{Zenith angle dependence of the intensity of showers
having constant ($\log_{10}N_\mu,\log_{10}N_e$)=(5.25-5.45,6.8-7.0)
for different primary compositions. Showers were simulated
with SIBYLL 2.1. The points are joined by straight
lines to guide the eye. To avoid overlapping of the results
for different compositions, they are multiplied by different
arbitrary factors.
}
\label{zenith_compos}
\end{figure}

 The composition we chose in this analysis is 
 completely ``ad-hoc''. In particular we have assumed an
 energy-independent ratio of the different elemental
 contributions. However, we have repeated the analysis for
 many different combinations of primary fractions, and only
 those with a large fraction of iron produce an exponential
 decrease of the intensity of showers with zenith angle.
 This is shown in Fig.~\ref{zenith_compos}. 

 We have not
 attempted to perform a fit to the intensity versus zenith angle 
 using the different fractions of primaries as parameters in the fit.
 Such an analysis would require the use of a true detector
 Monte Carlo simulation and could only be performed by the
 experimental group.
 However from the combinations we have experimented with we
 conclude that at least 60-70\% of iron is needed to produce a
 straight line in the nominal $N_\mu, N_e$bin. The Akeno
 measurements then strongly indicate that a large fraction of
 iron is present in the cosmic ray spectrum in the energy region
 between $10^{16}-10^{17}$ eV. This result is in qualitative
 agreement with recent analyses of the region around and above
 the knee in the cosmic ray spectrum (for example, KASCADE \cite{Ulrich01}
 and HiRes-MIA \cite{AZayyad01} measurements, see also \cite{Swordy02a}).
Finally, we note that an early analysis~\cite{TKGNature} using the method
of constant intensity cuts reached a similar conclusion about heavy
composition in this energy range based on data from the
BASJE air shower experiment on Mt. Chacaltaya~\cite{Bradtetal65}.

\section{\label{conclusions} Discussion and conclusions}

  We have investigated the influence of air shower fluctuations on
 the widely used constant intensity cut method. We consider two types
 of applications: the classic integral approach to the derivation
 of the cosmic ray energy spectrum, and the differential
 $N_\mu, N_e$ cut used for the derivation of the proton-air cross section.

  We find that the constant intensity cuts method can work for
 comparisons of data taken at different atmospheric depths and
 different angles. This is however possible only when the
 chemical composition of the primary cosmic rays is well known.
 The use of incorrect chemical composition can lead to a shift
 in the normalization of the energy spectrum. In the case of
 energy dependent composition the normalization errors for
 different cosmic ray flux components could also affect
 significantly the derived spectral index $\gamma$. Such shifts
 are also possible close to the detector $N_e$ threshold, where
 measurements at different zenith angles would detect showers
 of different composition. The larger the zenith angle $\theta$,
 the lighter would be the composition of the detected showers.


 The influence of shower fluctuations is much bigger when
 the constant intensity cut is used in a differential way
 to compare showers with same electron and muon sizes detected
 at different zenith angles.  The selection by the muon size
 $N_\mu$ with constant intensity cuts is indeed a very good
 method and leads to a good angle independent energy selection.
 This is the result of the much slower absorbtion of the shower
 muons as well as to the smaller $N_\mu$ fluctuations in showers
 with fixed primary energy.


 The constant $N_e-N_\mu$ method, which is used for derivation
 of the proton-air production cross section, is dominated by
 fluctuations even in the case of a pure proton composition.
 The accuracy of this method improves with the selection of
 showers with large $N_e$ for a fixed $N_\mu$ bin, where an 
 experiment would run out of statistics.  A possible improvement 
 of the method would be to use Monte Carlo shower simulations
 to determine zenith angle dependent $N_e$ bins. This, however,
 would represent a new method which is very different from the
 original idea of constant intensity cuts.

 The Akeno data that were used for the derivation of the proton-air
 production cross section can be interpreted in terms of cosmic
 ray composition. The angle independent exponential slope of the
 shower absorption length indicates a substantial fraction of heavy
 primaries in the energy range of 10$^{16}$ - 10$^{17}$ eV. 
 The comparison to showers simulated with the QGSjet98 and SIBYLL 2.1 
 hadronic interactions models require 60-70\% of iron in the primary
 cosmic ray flux to explain the absorption length derived by the
 Akeno group.

 These conclusions depend only mildly on the hadronic interaction
 model used in the simulation.

\noindent
{\bf Acknowledgments}
We thank A.A. Watson for helpful discussions.
J.A. Ortiz is supported by CAPES ``Bolsista da CAPES -
Bras\'{\i}lia/Brasil'' and acknowledges Bartol Research
Institute for its hospitality. This research is supported
in part by NASA Grant NAG5-10919.
RE, TKG \& TS are also supported by the US Department of Energy contract
DE-FG02 91ER 40626. The simulations presented here were performed on
Beowulf clusters funded by NSF grant ATM-9977692.

\bibliography{xsection}

\begin{thebibliography}{22}
\expandafter\ifx\csname natexlab\endcsname\relax\def\natexlab#1{#1}\fi
\expandafter\ifx\csname bibnamefont\endcsname\relax
  \def\bibnamefont#1{#1}\fi
\expandafter\ifx\csname bibfnamefont\endcsname\relax
  \def\bibfnamefont#1{#1}\fi
\expandafter\ifx\csname citenamefont\endcsname\relax
  \def\citenamefont#1{#1}\fi
\expandafter\ifx\csname url\endcsname\relax
  \def\url#1{\texttt{#1}}\fi
\expandafter\ifx\csname urlprefix\endcsname\relax\def\urlprefix{URL }\fi
\providecommand{\bibinfo}[2]{#2}
\providecommand{\eprint}[2][]{\url{#2}}

\bibitem[{\citenamefont{Alvarez-Mu\~niz et~al.}()}]{Alvarez-Muniz02c}
\bibinfo{author}{\bibfnamefont{J.}~\bibnamefont{Alvarez-Mu\~niz}}
  \bibnamefont{et~al.}, \bibinfo{note}{in preparation}.

\bibitem[{\citenamefont{Hara et~al.}(1983)}]{Hara83}
\bibinfo{author}{\bibfnamefont{T.}~\bibnamefont{Hara}} \bibnamefont{et~al.},
  \bibinfo{journal}{Phys. Rev. Lett.} \textbf{\bibinfo{volume}{50}},
  \bibinfo{pages}{2058} (\bibinfo{year}{1983}).

\bibitem[{\citenamefont{Baltrusaitis et~al.}(1984)}]{Baltrusaitis84}
\bibinfo{author}{\bibfnamefont{R.~M.} \bibnamefont{Baltrusaitis}}
  \bibnamefont{et~al.}, \bibinfo{journal}{Phys. Rev. Lett.}
  \textbf{\bibinfo{volume}{52}}, \bibinfo{pages}{1380} (\bibinfo{year}{1984}).

\bibitem[{\citenamefont{Honda et~al.}(1993)}]{Honda93}
\bibinfo{author}{\bibfnamefont{M.}~\bibnamefont{Honda}} \bibnamefont{et~al.},
  \bibinfo{journal}{Phys. Rev. Lett.} \textbf{\bibinfo{volume}{70}},
  \bibinfo{pages}{525} (\bibinfo{year}{1993}).

\bibitem[{\citenamefont{Aglietta et~al.}({\natexlab{a}})}]{Aglietta97}
\bibinfo{author}{\bibfnamefont{M.}~\bibnamefont{Aglietta}}
  \bibnamefont{et~al.}, \bibinfo{note}{in {\it Proceedings of the 25th
  International Cosmic Ray Conference}, Durban, South Africa, 1997, (World
  Scientific, Singapore, 1997), Vol. 6, p. 37.}

\bibitem[{\citenamefont{Aglietta et~al.}({\natexlab{b}})}]{Aglietta99}
\bibinfo{author}{\bibfnamefont{M.}~\bibnamefont{Aglietta}}
  \bibnamefont{et~al.}, \bibinfo{note}{in {\it Proceedings of the 26th Int'l
  Cosmic Ray Conference}, Salt Lake City, Utah, USA, 1999, (AIP, Melville, NY,
  2000), Vol. 1, p. 143.}

\bibitem[{\citenamefont{Aglietta et~al.}(1999)}]{Aglietta99b}
\bibinfo{author}{\bibfnamefont{M.}~\bibnamefont{Aglietta}}
  \bibnamefont{et~al.}, \bibinfo{journal}{Nucl. Phys. B (Proc. Suppl.)}
  \textbf{\bibinfo{volume}{75A}}, \bibinfo{pages}{222} (\bibinfo{year}{1999}).

\bibitem[{\citenamefont{Ellsworth et~al.}(1982)}]{Elsworthetal82}
\bibinfo{author}{\bibfnamefont{R.~W.} \bibnamefont{Ellsworth}}
  \bibnamefont{et~al.}, \bibinfo{journal}{Phys. Rev.}
  \textbf{\bibinfo{volume}{D26}}, \bibinfo{pages}{336} (\bibinfo{year}{1982}).

\bibitem[{\citenamefont{Bradt et~al.}()}]{Bradtetal65}
\bibinfo{author}{\bibfnamefont{H.}~\bibnamefont{Bradt}} \bibnamefont{et~al.},
  \bibinfo{note}{in {\it Proceedings of the 9th International Cosmic Ray
  Conference}, London, 1965, (The Institute of Physics and the Physical
  Society, London, 1965), Vol. 2, p. 715.}

\bibitem[{\citenamefont{Ave et~al.}(2001)\citenamefont{Ave, Knapp, Lloyd-Evans,
  Marchesini, and Watson}}]{Ave01a}
\bibinfo{author}{\bibfnamefont{M.}~\bibnamefont{Ave}},
  \bibinfo{author}{\bibfnamefont{J.}~\bibnamefont{Knapp}},
  \bibinfo{author}{\bibfnamefont{J.}~\bibnamefont{Lloyd-Evans}},
  \bibinfo{author}{\bibfnamefont{M.}~\bibnamefont{Marchesini}},
  \bibnamefont{and} \bibinfo{author}{\bibfnamefont{A.~A.}
  \bibnamefont{Watson}}, \bibinfo{journal}{Astroparticle Physics in press}
  (\bibinfo{year}{2001}), \eprint{astro-ph/0112253}.

\bibitem[{\citenamefont{Alvarez-Mu\~niz
  et~al.}(2002)\citenamefont{Alvarez-Mu\~niz, Engel, Gaisser, Ortiz, and
  Stanev}}]{Alvarez02}
\bibinfo{author}{\bibfnamefont{J.}~\bibnamefont{Alvarez-Mu\~niz}},
  \bibinfo{author}{\bibfnamefont{R.}~\bibnamefont{Engel}},
  \bibinfo{author}{\bibfnamefont{T.~K.} \bibnamefont{Gaisser}},
  \bibinfo{author}{\bibfnamefont{J.~A.} \bibnamefont{Ortiz}}, \bibnamefont{and}
  \bibinfo{author}{\bibfnamefont{T.}~\bibnamefont{Stanev}},
  \bibinfo{journal}{Phys. Rev. D in press}  (\bibinfo{year}{2002}).

\bibitem[{\citenamefont{Nagano and Watson}(2000)}]{NaganoWatson00}
\bibinfo{author}{\bibfnamefont{M.}~\bibnamefont{Nagano}} \bibnamefont{and}
  \bibinfo{author}{\bibfnamefont{A.~A.} \bibnamefont{Watson}},
  \bibinfo{journal}{Rev. of Mod. Phys} \textbf{\bibinfo{volume}{72}},
  \bibinfo{pages}{689} (\bibinfo{year}{2000}).

\bibitem[{\citenamefont{Engel et~al.}({\natexlab{a}})\citenamefont{Engel,
  Gaisser, Lipari, and Stanev}}]{Engel99}
\bibinfo{author}{\bibfnamefont{R.}~\bibnamefont{Engel}},
  \bibinfo{author}{\bibfnamefont{T.~K.} \bibnamefont{Gaisser}},
  \bibinfo{author}{\bibfnamefont{P.}~\bibnamefont{Lipari}}, \bibnamefont{and}
  \bibinfo{author}{\bibfnamefont{T.}~\bibnamefont{Stanev}}, \bibinfo{note}{in
  {\it Proceedings of the 26th Int'l Cosmic Ray Conference}, Salt Lake City,
  Utah, USA, 1999, (AIP, Melville, NY, 2000), Vol. 1, p. 415.}

\bibitem[{\citenamefont{Engel et~al.}({\natexlab{b}})\citenamefont{Engel,
  Gaisser, and Stanev}}]{Engel01}
\bibinfo{author}{\bibfnamefont{R.}~\bibnamefont{Engel}},
  \bibinfo{author}{\bibfnamefont{T.~K.} \bibnamefont{Gaisser}},
  \bibnamefont{and} \bibinfo{author}{\bibfnamefont{T.}~\bibnamefont{Stanev}},
  \bibinfo{note}{in {\it Proceedings of the 27th International Cosmic Ray
  Conference}, Hamburg, Germany, 2001, (Copernicus Gesellschaft,
  Katlemburg-Lindau, 2001), p. 431.}

\bibitem[{\citenamefont{Kalmykov et~al.}(1997)\citenamefont{Kalmykov,
  Ostapchenko, and Pavlov}}]{qgsjet}
\bibinfo{author}{\bibfnamefont{N.~N.} \bibnamefont{Kalmykov}},
  \bibinfo{author}{\bibfnamefont{S.~S.} \bibnamefont{Ostapchenko}},
  \bibnamefont{and} \bibinfo{author}{\bibfnamefont{A.~I.}
  \bibnamefont{Pavlov}}, \bibinfo{journal}{Nucl. Phys. B (Proc. Suppl.)}
  \textbf{\bibinfo{volume}{52B}}, \bibinfo{pages}{17} (\bibinfo{year}{1997}).

\bibitem[{\citenamefont{Nagano et~al.}(1984)}]{Nagano84}
\bibinfo{author}{\bibfnamefont{M.}~\bibnamefont{Nagano}} \bibnamefont{et~al.},
  \bibinfo{journal}{J. Phys. G} \textbf{\bibinfo{volume}{10}},
  \bibinfo{pages}{1295} (\bibinfo{year}{1984}).

\bibitem[{\citenamefont{Ulrich et~al.}()}]{Ulrich01}
\bibinfo{author}{\bibfnamefont{H.}~\bibnamefont{Ulrich}} \bibnamefont{et~al.},
  \bibinfo{note}{in {\it Proceedings of the 27th International Cosmic Ray
  Conference}, Hamburg, Germany, 2001, (Copernicus Gesellschaft,
  Katlemburg-Lindau, 2001), p. 97.}

\bibitem[{\citenamefont{Abu-Zayyad et~al.}(2001)}]{AZayyad01}
\bibinfo{author}{\bibfnamefont{T.}~\bibnamefont{Abu-Zayyad}}
  \bibnamefont{et~al.}, \bibinfo{journal}{Astropart. Phys.}
  \textbf{\bibinfo{volume}{16}}, \bibinfo{pages}{1} (\bibinfo{year}{2001}).

\bibitem[{\citenamefont{Swordy et~al.}(2000)}]{Swordy02a}
\bibinfo{author}{\bibfnamefont{S.~P.} \bibnamefont{Swordy}}
  \bibnamefont{et~al.}, \bibinfo{journal}{Astropart. Phys. in press}
  (\bibinfo{year}{2000}).

\bibitem[{\citenamefont{Gaisser}(1974)}]{TKGNature}
\bibinfo{author}{\bibfnamefont{T.}~\bibnamefont{Gaisser}},
  \bibinfo{journal}{Nature} \textbf{\bibinfo{volume}{248}},
  \bibinfo{pages}{122} (\bibinfo{year}{1974}).

\bibitem[{\citenamefont{Nikolaev}(1993)}]{Nikolaev93b}
\bibinfo{author}{\bibfnamefont{N.~N.} \bibnamefont{Nikolaev}},
  \bibinfo{journal}{Phys. Rev.} \textbf{\bibinfo{volume}{D48}},
  \bibinfo{pages}{R1904} (\bibinfo{year}{1993}).

\bibitem[{\citenamefont{Engel et~al.}(1998)\citenamefont{Engel, Gaisser,
  Lipari, and Stanev}}]{Engel98c}
\bibinfo{author}{\bibfnamefont{R.}~\bibnamefont{Engel}},
  \bibinfo{author}{\bibfnamefont{T.~K.} \bibnamefont{Gaisser}},
  \bibinfo{author}{\bibfnamefont{P.}~\bibnamefont{Lipari}}, \bibnamefont{and}
  \bibinfo{author}{\bibfnamefont{T.}~\bibnamefont{Stanev}},
  \bibinfo{journal}{Phys. Rev.} \textbf{\bibinfo{volume}{D58}},
  \bibinfo{pages}{014019} (\bibinfo{year}{1998}).

\end{thebibliography}

\end{document}